\RequirePackage{fix-cm}
\documentclass[twocolumn]{svjour3}          
\smartqed  
\usepackage{graphicx}
\usepackage{natbib}
\usepackage{amsmath}
\usepackage[hyphens]{url}
\usepackage[ruled,vlined]{algorithm2e}
\usepackage{multirow}
\usepackage{tabularx}
\usepackage{soul}
\usepackage{tikz}
\usepackage{booktabs}
\usepackage{hyperref}
\usepackage{xcolor}

\journalname{Specialized Cyber Threat Intelligence}
\begin{document}\sloppy

\title{Multi-Level Fine-Tuning, Data Augmentation, and Few-Shot Learning for Specialized Cyber Threat Intelligence
}

\titlerunning{Specialized Cyber Threat Intelligence} 

\author{Markus Bayer \and Tobias Frey \and Christian Reuter
}


\institute{Markus Bayer \and Tobias Frey \and Christian Reuter \at
              Technical University of Darmstadt, Darmstadt, Germany \\
              \email{\{bayer, frey, reuter\}@peasec.tu-darmstadt.de}          
}

\date{bayer@peasec.tu-darmstadt.de \and tobiasjonathan.frey@stud.tu-darmstadt.de \and reuter@peasec.tu-darmstadt.de \\ PEASEC - Science and Technology for Peace and Security \\ Technical University of Darmstadt  \\ Pankratiusstraße 2, 64289 Darmstadt \\ Germany}

\twocolumn[
    \maketitle
    \begin{@twocolumnfalse}
        \begin{abstract}
        Gathering cyber threat intelligence from open sources is becoming increasingly important for maintaining and achieving a high level of security as systems become larger and more complex.
        However, these open sources are often subject to information overload. It is therefore useful to apply machine learning models that condense the amount of information to what is necessary.
        Yet, previous studies and applications have shown that existing classifiers are not able to extract specific information about emerging cybersecurity events due to their low generalization ability.
        Therefore, we propose a system to overcome this problem by training a new classifier for each new incident. Since this requires a lot of labelled data using standard training methods, we combine three different low-data regime techniques -- transfer learning, data augmentation, and few-shot learning -- to train a high-quality classifier from very few labelled instances. We evaluated our approach using a novel dataset derived from the Microsoft Exchange Server data breach of 2021 which was labelled by three experts.
        Our findings reveal an increase in F1 score of more than 21 points compared to standard training methods and more than 18 points compared to a state-of-the-art method in few-shot learning. Furthermore, the classifier trained with this method and 32 instances is only less than 5 F1 score points worse than a classifier trained with 1800 instances.

        \end{abstract}
    \end{@twocolumnfalse}
]

\section{Introduction} \label{introduction}
Social media is where cutting-edge and critical cyber threat information is disseminated, which is highly relevant to researchers, security providers, security operation centers, urban infrastructures, and cyber emergency response teams (CERTs), among others \citep{mittal2016cybertwitter,rodriguez2019generating}. 
While there have been several research works on general cyber threat detection \citep{dionisioEndtoendCyberthreatDetection2020, fangDetectingCyberThreat2020a}, the goal of this work is to enable a fine-grained information collection.

One major challenge in collecting specific information in this domain is that cyber information is highly dynamic and differs greatly from past events (in terms of specific names, different attack vectors, specific attack paths, affected functions, etc.) \citep{CHATTERJEE2020106664}. 
As a result, supervised machine learning yields poor results because these dynamics cannot be captured in the learning process. 
Alternatively, new classifiers could be trained for each cyber threat so that the new features are taken into account. 
However, since machine learning usually requires a large amount of data for normal training, this would result in having to label a dataset for each cyber threat, which is unrealistic considering the effort involved and the need for fast and up-to-date information. 
Against this background, the concept of active learning systems take a first step towards label reduction for supervised machine learning for cyber threats \citep{riebeCySecAlertAlertGeneration2021}.
Active learning supports the labeling process, so that only the instances with the highest learning value need to be labeled for machine learning.
However, despite this method, too much data is still needed to train a useful classifier.
The endeavor sought in this work takes an even stronger stance on labeling reduction by proposing a system with few-shot learning, transfer learning, and data augmentation. 
With few-shot learning, it is sufficient if the model is already trained with very few instances, as opposed to hundreds or thousands in the case of active or normal learning \citep{Brown2020}. 
This includes special learning techniques as well as transfer learning, where knowledge from a previous task is transferred to the new one.
Data augmentation is used to create artificial instances from the training data using label-preserving transformations \citep{bayerSurveyDataAugmentation2022}. 

The concept of few-shot learning is extended in this work through the use of multi-level transfer learning. 
The different levels start with a model that has been trained on a large general dataset and thus has a basic prior knowledge. 
During the next steps, this model is approximated more and more to the actual task domain. 
In this way, it can be ensured that the model is given a basic cybersecurity reference in order to be able to counter the dynamics in the task, in addition to being familiar with the task. 
This is particularly relevant for urban infrastructures, which require high resilience against cyberattacks, as well as for CERTs, as they need to collect and communicate information in the most reliable and targeted way possible \citep{riebeImpactOrganizationalStructure2021}. 
The data augmentation strategy is inspired by the work of \citet{bayerDataAugmentationNatural2021} and follows the example of \citet{yooGPT3MixLeveragingLargescale2021} by utilizing the large generation model GPT-3 to generate new instances based on the few existing labeled ones.

Our paper includes several contributions relevant for the cybersecurity and machine learning community:
\begin{itemize}
    \item A novel pipeline combining transfer learning, data augmentation, and few-shot learning for developing an effective specialized cyber threat intelligence (CTI) classifier.
    \item Novel techniques of data augmentation and few-shot learning to deal with a small number of training instances.
    \item A new specialized CTI dataset annotated by three experts and based on the 2021 Microsoft Exchange Server data breach. 
\end{itemize}

The code and dataset of this study are freely available. The remainder of the paper is structured as follows: After introducing related work on transfer learning, data augmentation, few-shot learning, and cyber threat detection and intelligence (Section \ref{related_work}), we explain the concept of our method (Section \ref{concept}).
It is subdivided in three components which are described in detail. 
In Section \ref{evaluation} the evaluation is presented and findings are given in detail. 
The last section (Section \ref{discussion}) contains a discussion of the implications, limitations, and potentials for future research.
\section{Related Work} \label{related_work}
\subsection{Transfer Learning} \label{transferlearning}
Transfer learning describes the process of transferring knowledge gained from training a neural network from one task to another related task  \citep{torrey2010transfer,pan2020transfer}. 
This technique is now one of the standard learning methods for machine learning, especially in the field of natural language processing (NLP).
It is particularly powerful for tasks where there is not enough training data or it is difficult to manually adjust the data for training. 
In this case, it is possible to use a pre-trained neural network that was trained to solve a related task or with more easily accessible data. 
Afterwards the neural network is fine-tuned with the task-specific data to fit the wanted task. 
One of the most frequently used pre-trained models is BERT by \cite{Devlin2018}. 
BERT (short for Bidirectional Encoder Representations from Transformers) is a pre-trained deep bidirectional transformer for language understanding. 
In essence, it is trained by predicting words in a sentence given the other words, also called masked language modeling. 
It has a lot of widely used descendants trained for many different tasks, such as BioBERT \citep{leeBioBERTPretrainedBiomedical2019}, SciBERT \citep{beltagySciBERTPretrainedLanguage2019}, and CamemBERT \citep{martin-etal-2020-camembert}. 
While BERT is already a considerably large model, nowadays far larger models, like GPT-3 from \cite{Brown2020}, are trained. 
Compared to BERT's base model with 110 million parameters, GPT-3 has 175 billion parameters, however, GPT-3 is not publicly available and cannot be easily fine-tuned due to its size.

\subsection{Data Augmentation}
Data augmentation is the concept for artificially enlarging the training datasets for machine learning by transforming the existing ones. 
Originated and heavily used in computer vision, it is now also increasingly being explored on textual data \citep{bayerSurveyDataAugmentation2022}. 
NLP data augmentation techniques can be applied to the raw text or also on the numerical representations. 
Ranging from small transformations, i.e. flipping characters \citep{Belinkov2018} or inducing adversarial noise \citep{jiangSMARTRobustEfficient2020}, to interpolated \citep{sunMixuptransfomerDynamicData2020} or even newly created instances \citep{Anaby-Tavor2020}, data augmentation can have great effects. 
Nevertheless, as \citet{Longpre2020} point out, the success of data augmentation in NLP is often not perceivable when fine-tuning large pre-trained models. 
A data augmentation technique needs to incorporate new linguistic patterns as otherwise the changes are too small and already captured by the pre-training phase of the model. 
For example, simple synonym replacement methods have not been shown to be beneficial with pre-trained models, as these synonyms are already mapped to nearly the same vector for their numerical representation \citep{mosolovaTextAugmentationNeural2018}. 
On the other hand, there are generation models that can integrate new linguistic patterns, for example, through their own training data during pre-training, as for example shown by \citet{yooGPT3MixLeveragingLargescale2021} with the GPT-3 model. 
The challenge with using these models is to make the generations truly label preserving. 
This is, for example, done by \citet{Anaby-Tavor2020}, \citet{queirozabonizioPretrainedDataAugmentation2020} and \citet{bayerDataAugmentationNatural2021}. 
The models are conditioned by fine-tuning on the label-induced training data (or just the class data) and are then tasked to complete a text given the label conditioned beginning (prompt). 
As this is oftentimes not sufficient to achieve a high label preservation, a filter mechanism is used that removes artificial instances that are unlikely to fit the class.
For example, \citet{Anaby-Tavor2020} use a classifier trained on the data to predict whether the new instance can be assigned to the expected label. 

For an overview of the data augmentation methods that could be used in this study, we advise the reader to have a look at the survey from \cite{bayerSurveyDataAugmentation2022}.

\subsection{Few-Shot Learning} \label{realtedwork_fewshot}
Few-shot learning describes the training of effective classifiers on the basis of a small number of examples. 
While there are several strands of research on few-shot learning \citep{braggFLEXUnifyingEvaluation2021}, in this study we focus on the use of pre-trained language models. 
At the latest, the large language model GPT-3 by \citet{Brown2020} paved the way for using these kinds of models, as it reaches astounding performance even without task-specific training data. 
However, as GPT-3 is too large for most companies and research institutes, the research field adapted smaller language models to reach similar or even better few-shot performances \citep{tamImprovingSimplifyingPattern2021}. 

Pre-trained language models can be especially beneficial for few-shot settings when the instances are reformulated in a cloze-style way.
Cloze tests \citep{taylorClozeProcedureNew1953} are tests where some words in the text are missing and have to be completed. 
For few-shot learning, instances are rephrased, often into questions, so that the text contains the label (or a word that can be mapped to the label), generally within the answer to the question. The label, known (training) or not known (testing and inference), is masked out, so that the language model can fill it with the right word and a label can be inferred. 
Using the language model directly is more effective for few-shot learning than the classical way of training a classifier head on top of it, as there are no more randomly initialized parameters that have to be learned \citep{gaoMakingPretrainedLanguage2021}. 

\begin{figure*}
\caption{Example of a template and a verbalizer and how they are applied on an instance.}
\label{template}
\begin{center}
\includegraphics[width={0.9\textwidth}]{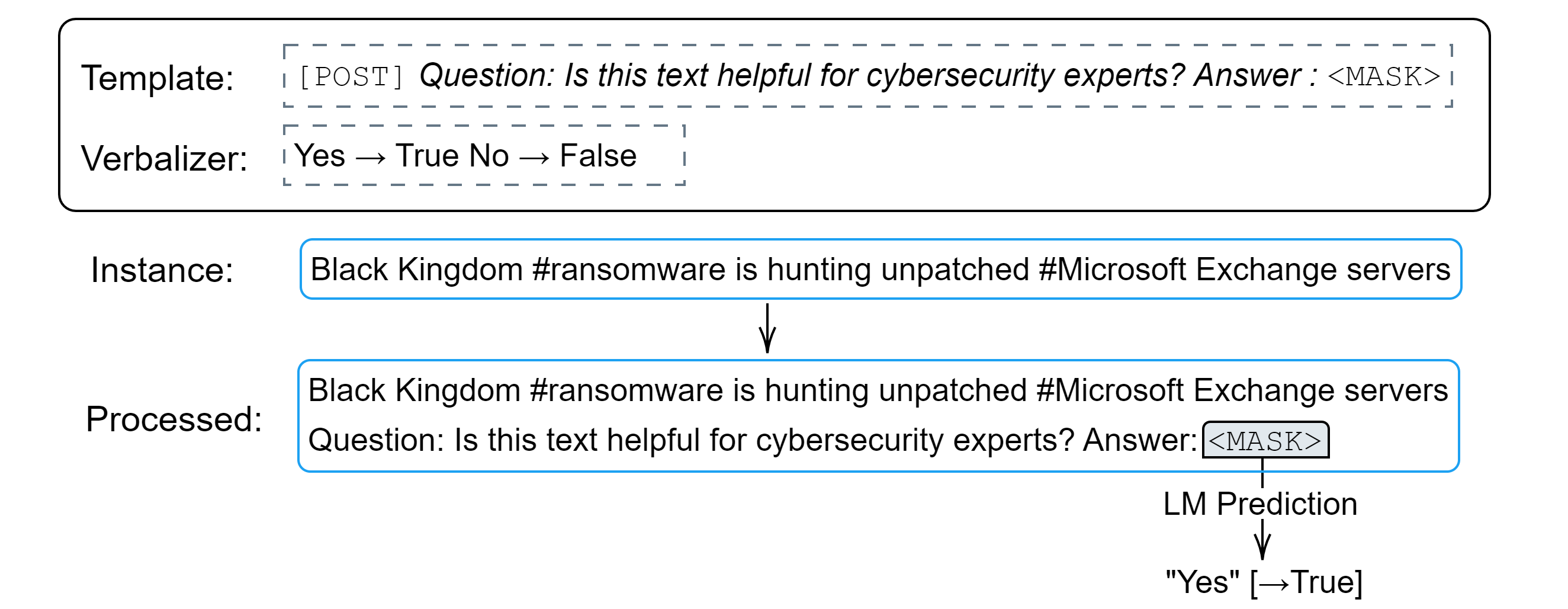}
\end{center}
\end{figure*}

A pattern describes the transformation of the input instance to the cloze-like text.
The verbalizer maps the predicted words for the mask to the label. 
An example for a pattern and a verbalizer can be seen in Figure \ref{template}.

\citet{gaoMakingPretrainedLanguage2021} show that the choice of template and verbalizer has a major impact on the resulting performance. 
Since domain knowledge is often necessary for these, the authors propose a method to automatically find meaningful templates and verbalizer. 
For this purpose, they use a language model and the existing training instances to predict the words for the verbalizer and template.
\citet{zhangDifferentiablePromptMakes2022} take a different perspective on automatic template generation with the DART method by making the template differentiable. 
They use special tokens in the template that are mapped into trainable parameters.  These template parameters are then optimized together with
the target label. 
PERFECT by \citet{mahabadiPERFECTPromptfreeEfficient2022} leverages task-specific adapters to replace template tokens. Adapters make it possible to train only the newly added parameters, which are able to transform the hidden states, while freezing all other parameters.

\citet{schickExploitingClozeQuestions2021} propose a semi-supervised few-shot learning technique, called PET.
They take several manually designed templates and use the training data to train on each one a pre-trained language model. 
They take these models to generate pseudo-labels for unlabeled data. 
A classifier is then trained on the resulting dataset.
\citet{tamImprovingSimplifyingPattern2021} adapt the PET method to not be dependent on additional training data and can even improve the performances. 
Contrary to the preceding PET technique, the word probabilities are computed not only for the verbalizer words, like ``yes'' and ``no'', but also for all other words. 
In the training, incorrect class tokens are explicitly penalized and correct tokens are encouraged. 
Furthermore, ADAPET \citep{tamImprovingSimplifyingPattern2021} introduces a label conditioning step in which the model is tasked to predict other tokens in the sentence given the label.

\subsection{Cyber Threat Detection and Intelligence}

Cyber threat detection is generally known as the process of automatic scraping of the webspace and Open Source Intelligence (OSINT) to detect possible cybersecurity vulnerabilities \citep{Sabottke2015, riebeCySecAlertAlertGeneration2021, Sceller2017}. 
Social Media platforms, like Twitter, are part of OSINT and propose a great space to share and discuss possible cybersecurity vulnerabilities. 
There are some automated systems and research that already scrape Twitter and other OSINT sources to detect cyber threats. 
Some examples are the \mbox{\textit{CySecAlert}} system from \cite{riebeCySecAlertAlertGeneration2021} or \mbox{\textit{SONAR}} from \cite{Sceller2017}, which collect cyber threat relevant tweets from Twitter, filter them, and present them in a manageable dashboard.

CTI on the other hand describes the process of collecting additional information after the first detection of a cyber threat. 
The process helps deliver the context of the vulnerabilities found to assist CERTs and cybersecurity organizations make sound decisions and find quick solutions \citep{abu2018cyber, TOUNSI2018212, WAGNER2019101589}. CTI is currently mostly accomplished by manually sharing information on different platforms \citep{abu2018cyber}. 
They depend heavily on manual input and are therefore labor intensive \citep{WAGNER2019101589}. 
However, there are already some threat intelligence platforms, such as Facebook ThreatExchange or CrowdStrike, that are able to automatically detect, monitor, and analyze cyber threat occurrences \citep{TOUNSI2018212}.
A manageable dashboard is also provided by the Cyber Threat Observatory of \cite{kaufholdCyberThreatObservatory2022}, which aggregates cybersecurity information from various sources, including social media, security advisories, indicators of compromise and CVEs. 
However, these systems need too much time to adapt to a newly discovered threat that is, for example, propagated on Twitter.

\subsection{Research Gap} \label{research_gap}

Our study addresses several research gaps which are highly relevant for researchers as well as practitioners. 
Most importantly, our research paves the way for fine-grained CTI.
Current research addresses CTI from a very coarse perspective, by building classifiers, like \citet{riebeCySecAlertAlertGeneration2021}, that are able to find general cyber threat information. 
As a result, only a small amount of data reduction can be achieved in these information-overloaded situations. 
On the other hand, specialized classifiers are not designed to generalize well to new situations. 
Our work fills this gap by introducing a pipeline for specialized CTI, where new cyber threat events are encountered with the very fast creation of new classifiers. 
By addressing this fine-grained information gathering challenge, we create a novel dataset combined with a sophisticated labeling guideline for CTI. 
Furthermore, with our pipeline we address research gaps of machine learning low-data regimes. 
Our data augmentation strategy is the first to explore the generation capabilities of large language models with constraining them through filtering mechanisms. 
We combine the works of \citet{yooGPT3MixLeveragingLargescale2021} and \citet{bayerDataAugmentationNatural2021} by using GPT-3 with a human-in-the-loop filtering mechanism.
We extend the few-shot learning research by proposing a multi-level fine-tuning approach. 
In the process, the model learns a very broad knowledge in the first levels, which in the later stages becomes more and more directed to the specific CTI task.
\section{Concept} \label{concept}
\subsection{Dataset Creation} \label{dataset_creation}

The goal of dataset creation is to extract specific CTI information during a significant cyber threat event. 
This dataset is subsequently binary-labeled according to the relevance of the information for CTI and for cybersecurity experts. 
We focused on the Microsoft Exchange Server data breach of 2021, where four zero-day exploits were discovered.
While the first report of a vulnerability was already made in January of that year, in March various attackers were found to be exploiting the vulnerabilities and a Proof of Concept was released. 

We used the Twitter APIv2 to gather Tweets in March that fulfill the query \emph{``Microsoft Exchange'' OR ``MS Exchange'' OR ``CVE-2021-26855'' OR ``CVE-2021-26857'' OR ``CVE-2021-26858'' OR ``CVE-2021-27065''}. 
For these Tweets we resolved the links that were shortened by Twitter, as the full URLs might be an important indicator in the context of CTI. 

The labeling process of the data was performed by three cybersecurity experts guided by a codebook. 
The guidelines, which provide clear guidance on when to mark a contribution as relevant or irrelevant, were updated iteratively by the annotation leader.
A first draft was developed using the CTI concept \citep{robmcmillanDefinitionThreatIntelligence2013}: 
\begin{quote}
    ``Threat intelligence is referred to as the task of gathering evidence-based knowledge, including context, mechanisms, indicators, implications, and actionable advice, about an existing or emerging menace or hazard to assets that can be used to inform decisions regarding the subject’s response to that menace or hazard.''
\end{quote}
After an initial sifting of the tweets and again after the first labeling of 750 tweets, the process was refined by the annotation leader. 

The first round of annotation of 750 tweets was conducted by the annotation leader, who updated the guidelines after gathering several insights. 
He and the other two cybersecurity experts then annotated the 750 tweets again. 
After this round, all three experts discussed the cases they were not sure about and corrected them if necessary. 
Regarding the intercoder reliability the Kappa Scores were calculated (see \ref{tab1:kappascore}). 
Subsequently, each annotator tagged 750 different examples, resulting in a total of 3001 commented Twitter posts for the complete dataset (the labels of the 750 instances of the first round were determined by majority vote). 

\begin{table}[]
    \centering
    \begin{tabular}{c|c}
         Coder & Score \\
         \hline
         C1 and C2 & 0.8763 \\
         C2 and C3 & 0.7446 \\
         C1 and C3 & 0.8709 
    \end{tabular}
    \caption{Intercoder reliability calculated with the Kappa Score.}
    \label{tab1:kappascore}
\end{table}

The dataset was then split into a full and few-shot training set and development set. The splits (train, dev) consist of 1800 and 600 instances for the full set and 32 and 32 instances for the few-shot set, respectively. 
The test set is the same in both cases and consists of 601 instances.

\subsection{Approach} \label{approach}
\begin{figure}

    \includegraphics[width=0.5\textwidth]{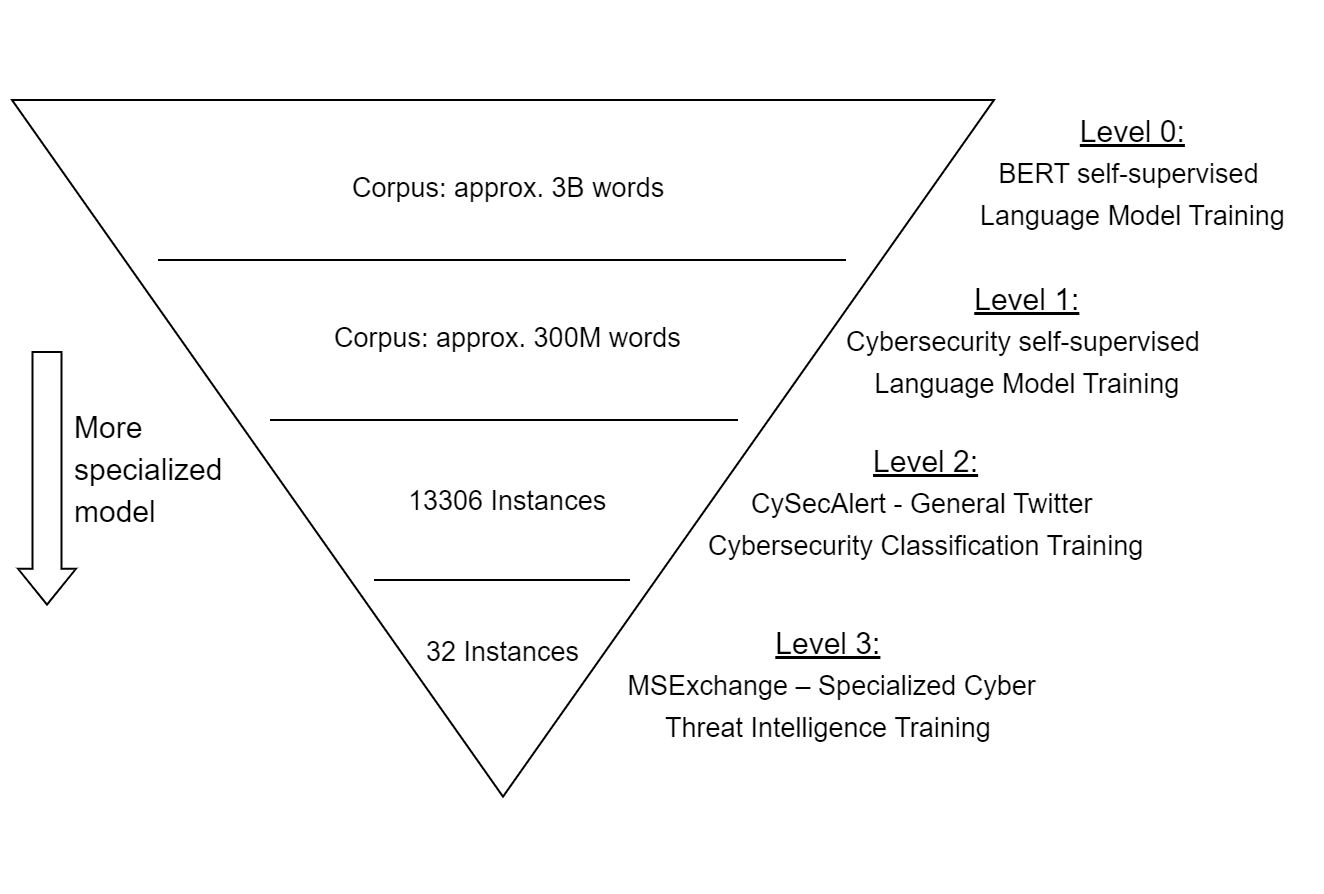}
    \caption{Multi-level fine-tuning process that shows the model becoming more specialized as it is guided to the actual task with less data.}
    \label{fig:mulitlevel}
\end{figure}

Our system for dynamic, specialized cyber threat detection consists of three components, all of which help to boost performance with little data. 
We explain the three components in detail in the following:

\paragraph{Multi-Level Fine-Tuning:} In light of the success of large pre-trained models such as BERT, we propose to further tune such models on several levels of domain-dependent data (see Figure \ref{fig:mulitlevel}). 
The levels begin from a broader view and are narrowed down to the actual task. 
In our case, we first take a pre-trained BERT model (which can be seen as the 0th level of fine-tuning), train it with masked language modeling on cybersecurity data.
We then tune the resulting model for classification on the CySecAlert dataset \citep{riebeCySecAlertAlertGeneration2021} in which Twitter posts are generally assigned to the cybersecurity domain. Finally, we train it on the few training examples of the specialized cyber threat dataset. 
The rationale behind this is that the model gains more and more knowledge as it is tuned to more and more fitting tasks. 
The 0th level is about gaining general knowledge of text. 
In the first level, the dataset consists of papers, blogs, web pages, and also Twitter data, from which the model gains knowledge about cybersecurity language and also how Twitter data is written in this domain. 
In the second level, the model should gain a general understanding of the relevance of cybersecurity information. 
Finally, in the third level, the model is tuned to the actual task data to which it can transfer the knowledge of the previous levels. 

\paragraph{GPT-3 Data Augmentation:} With data augmentation we can create new instances from existing ones, which can be particularly advantageous when the amount of data is small. 
We propose a data augmentation strategy based on text generation with GPT-3 \citep{Brown2020}, which is inspired by the method from \citet{yooGPT3MixLeveragingLargescale2021} and \citet{bayerDataAugmentationNatural2021}. 
GPT-3 can be tasked to complete a given text, also called a prompt. 
We utilize this mechanism so that the generation model creates new instances based on the training data of one class. 
Specifically, this means that we are concatenating all instances of one class with a class specific priming token. 
For the class of cyber threat information we prepend every positive instance with ``cybersecurity \verb|->|''.
For the irrelevant class we chose ``other \verb|->|'' as priming token. In both cases the priming token is also appended at the end so that the model generates the instance(s) after it. 
Dependent on how many remaining generation tokens the model has after the prompt, it may generate more than one instance by picking up the priming token. 
After the creation of the instances we perform the human-in-the-loop filtering step proposed by \citet{bayerDataAugmentationNatural2021}. The training examples and generated instances are mapped into an embedding space. 
There, the generated instances that deviate the most from the training data are discarded. 
The distance from which this happens is determined by an expert.

\paragraph{Few-Shot Learning:} We make use of the existing ADAPET \citep{tamImprovingSimplifyingPattern2021} few-shot learning technique and adapt it to our case. 
With ADAPET, in contrast to normal use, no classification head is trained on the language models. 
The instances are transformed to  cloze-style phrases and then the language model itself is used to predict the blank word in the phrases. 
The predicted word is subsequently transformed with a verbalizer to one of the labels. 
The cloze-style phrases are automatically formed with templates.
For our task we use the following template: 

\begin{quote}
``\texttt{[POST]}  Question : Is this text helpful for cybersecurity experts? Answer : \texttt{$<$MASK$>$}. \texttt{[SEP]}''
\end{quote}

The verbalizer maps the two possible words ``yes'' and ``no'' to the labels representing relevant and not relevant. 
As explained in Section \ref{realtedwork_fewshot} there also exist methods for automatically determining the pattern and verbalizer. 
We believe that these techniques are not necessary in our case, as we can integrate the expert knowledge regarding the task, which facilitates the learning process.
\section{Evaluation} \label{evaluation}

\subsection{Dataset, Models and Evaluation Settings}

Following the research goal of specialized CTI for security professionals, we constructed a setting, consisting of models and datasets, representing the real conditions. 
For the dataset, we labeled data from the 2021 Microsoft Exchange Server data breach. 
The specifics of the dataset can be found in Section \ref{dataset_creation}.
The labeled dataset, including few-shot and normal-shot splits, is freely available.

In our main evaluation we have different settings regarding the dataset and models.
The \textit{baseline} and initial model of our evaluation is the bert-base-uncased model by \citet{Devlin2018}. 
For the baseline, this model is fine-tuned on the few-shot dataset representing the standard training strategy without any few-shot or data augmentation methods.
For the \textit{best case}, on the other hand, we train the bert-base-uncased model with the full dataset of 1800 instances.
This is called the best case because we consider this amount of data to be the best case in the event of a new cybersecurity attack.
In addition, we also train a model with ADAPET, as we consider this to be the current state of the art in few-shot research. 
In preliminary tests, we found that ADAPET performed best on the few-shot split with ALBERT \citep{lanALBERTLiteBERT2020} compared to DART and a PERFECT variant.
To be consistent with our evaluation settings as opposed to the evaluation settings of ADAPET, we use the bert-base-uncased model, instead of the albert-xxlarge-v2 model by \citet{tamImprovingSimplifyingPattern2021}.
The evaluation settings of our procedure are divided into the three components mentioned. 
For the data augmentation technique we use GPT-3 (DaVinci) as text generation model, which is prompted with the specifics explained in section \ref{approach}.
The multi-stage fine-tuning process starts with the bert-base-uncased model, which is further pre-trained on a cybersecurity dataset, which is then fine-tuned with the ADAPET few-shot method on the CySecAlert dataset.
This resulting model is finally trained on the few-shot split and evaluated on the test set of the Microsoft Exchange dataset. 
Furthermore, in addition to the CySecAlert fine-tuning process, we also use the ADAPET few-shot method for the fine-tuning of the Microsoft Exchange Server dataset.
The mentioned components are also inspected within an ablation study, showing their individual contribution to the overall pipeline.

The evaluation performance is measured in accuracy and with the F1-score. 
For every evaluation setting, we perform five runs to rule out random factors.
The results are given with the minimum, maximum, mean, and standard deviation.


\subsection{Hyperparameters}

As already mentioned, we are using bert-base-uncased as base model for our experiments. 
The evaluations are performed on a NVIDIA A100 with 40 GB GPU memory.
The training runs on the CySecAlert and Microsoft Exchange dataset are performed with 5 epochs each.
Furthermore, we used a batch size of 48, 100 warmup steps with a warmup ratio of 0.06, a learning rate of 0.00001, and weight decay of  0.001.
As optimization algorithm, we used the Adam algorithm.
For the data augmentation technique we used the GPT-3 text-davinci-002, which has 175 billion parameters.
The filtering was performed with SBERT with the all-mpnet-base-v2 model.

\subsection{Evaluation}
The first section of our evaluation is about the data augmentation process, as we manually inspected the instances generated by GPT-3. 
After this, the main evaluation follows where we compare our methods to a baseline, state-of-the-art and best case experiment.
Finally, we inspect our method by doing ablation studies, testing how each component evaluates.

\subsubsection{Data Augmentation} \label{eval:dataaugmentation}

Due to our human-in-the-loop approach, we already saw that the generated instances are of very high quality. 
An excerpt of the generated data is given in Table \ref{table:generated_data}. 
For research purposes, we were also interested in the most likely original instances that the model used for generating specific instances. 
This is why we tried to find the training instance with the closest resemblance to the generated one. 
We measured the resemblance by generating sentence embeddings with SBERT \citep{reimersSentenceBERTSentenceEmbeddings2019} and comparing them with the cosine distance. 
These counterparts are also given in Table \ref{table:generated_data}. 
These examples show that the data augmentation method is capable of many different transformations. 
The first example demonstrates that the model sometimes replaces one or few words with synonyms (\textit{hosting}  \verb|->| \textit{running}) or adds context words (\textit{\#cybersecurity}). 
While in the second example, one can see that the model is able to paraphrase parts of the original instance (\textit{Another \#ransomware operation known as `Black Kingdom' is exploiting the [...]} \verb|->| \textit{Black  Kingdom  ransomware  is  exploiting  the [...]}), in the third example the entire instance is paraphrased (\textit{Just as predicted, the Microsoft Exchange exploit chain \#ProxyLogon now confirmed being used to install ransomware} \verb|->| \textit{The ProxyLogon vulnerability in Microsoft Exchange Server is being actively exploited in the wild to install ransomware}).
The fourth shown instance is an example of the method stripping away parts, while still preserving the label (\textit{\st{Thousands of US companies have been hacked by Chinese hackers using This RCE.} Microsoft Exchange Server Remote Code Execution CVE-2021-26855 Exploit.}).
For some generated instances, like the fifth example, we were not able to find similar instances. 
The instances might be entirely new based on the interpolation of the given instances and the knowledge of the underlying model.

Regarding the irrelevant class, we see that many generated instances are duplicates of the training instances, differing at most by very small changes, such as removing the hashtag in the first example (\textit{\#MicrosoftExchange Server Attack} \verb|->| \textit{Microsoft Exchange Server Attack}) or swapping the position of words in the second example (\textit{\#Technology \#TechNews Microsoft [...] Authority \#Cybersecurity \#AiUpNow \#techy} \verb|->| \textit{\#Technology \#Cybersecurity Microsoft [...] Authority \#AiUp-Now \#tech}). 
While the third example, again, shows an instance where the content is paraphrased, the last two generated texts have no clear counterpart.

\begin{table*}[t]
    \centering
    \begin{tabular}{l@{\hspace{1em}}l@{\hspace{1em}}l@{\hspace{1em}}l@{\hspace{1em}}}
    \toprule
    \textbf{Name}                    & \textbf{Model}     & \textbf{Accuracy}  & \textbf{F1}    \\ \midrule
    Best Case               & BERT      & {\scriptsize 84.69/} 85.36(0.07) {\scriptsize /86.02} & {\scriptsize 84.87/} 85.35(0.47) {\scriptsize /85.81}\\ 
    Baseline                & BERT      & {\scriptsize 46.26/} 49.65(1.90) {\scriptsize /50.58} & {\scriptsize 25.06/} 58.70(18.81) {\scriptsize /67.18} \\ 
    ADAPET                    & BERT &{\scriptsize 64.89/} 65.89(1.35) {\scriptsize /68.05} & {\scriptsize 59.30/} 62.54(4.32) {\scriptsize /69.81}                  \\ 
    Our Approach  & CyBERT    & {\scriptsize 78.54/} 79.13(0.56) {\scriptsize /80.03} & {\scriptsize 80.42/} 80.63(0.27) {\scriptsize /81.07} \\ 
    \bottomrule
    \end{tabular}
    \caption{Detailed evaluation results of the main experiments. The values on the left show the minimum, in the middle the mean, in brackets the standard deviation, and on the right the maximum value.}
    \label{tab:evaluation_results}
\end{table*}

\begin{figure}
    \centering
    \includegraphics[width=0.5\textwidth]{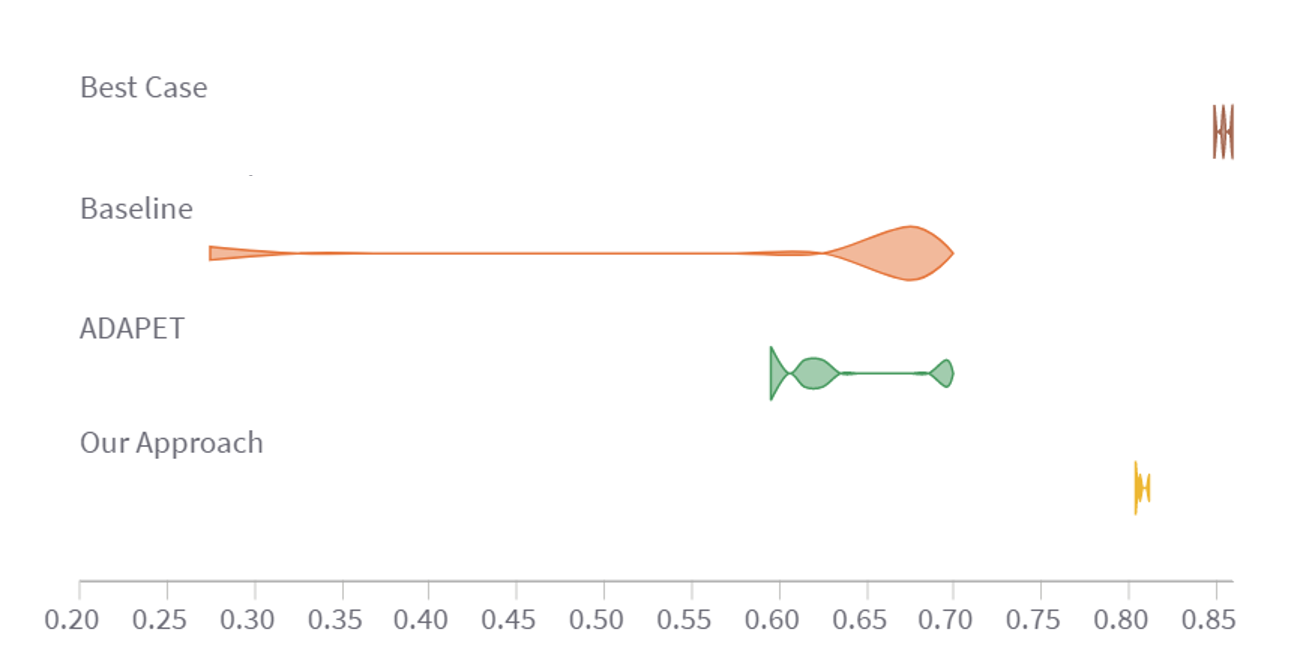}
    \caption{Violin plots of the main experimentation setting.}
    \label{fig:violinplots}
\end{figure}

\subsubsection{Main Experiments}

In our main experiments, we test the whole pipeline proposed in Section \ref{concept}. 
As a quick reminder, our method includes the multi-level fine-tuning with bert-base-uncased on cybersecurity data, the CySecAlert dataset and the actual few-shot learning task with 32 instances, as well as the GPT-3-based data augmentation technique and ADAPET for few-shot training.
For a sensible comparison, we first follow the standard training procedure by fine-tuning a bert-base-uncased model with a classifier head on the few-shot training instances (baseline).
Furthermore, we test a bert-base-uncased model with the ADAPET method, as it can be regarded as the state-of-the-art method for performing few-shot learning.
We also perform a best case evaluation in which we train the bert-base-uncased model on the full training dataset (1800 instances) to see how a classifier would perform with enough data.
A more detailed analysis of the approach itself can be found in the ablation studies in Section \ref{ablation}. 

The results of the pipeline experiments are shown in Table \ref{tab:evaluation_results}.
It is observable that the baseline is not able to learn any meaningful classification strategy with the low dataset, reaching an accuracy of about 50\% and F1 score of 58.70\%. 
ADAPET reaches a significantly higher accuracy with an additive improvement of about 15 points in accuracy and a F1 score of 62.57\%.
This is, nevertheless, far from a good classification quality as the best case classifier reaches a F1 score of 85.35\%. 
With an F1 score of 80.63\%, our approach proposed in this paper could even almost keep up with the best case classifier. 
Particularly noteworthy at this point is that the best case classifier is trained with 1800 instances, while our approach only has access to 32 instances. 
Furthermore, our approach improves the current state of the art with 18.09 points in F1.
A look at the violin plots in Figure \ref{fig:violinplots} shows that both the best case and our approach have a very good standard deviation, which means that both are robust to random changes.

The evaluation results show that our approach is able to identify cyber threat information from which we can deduce that a new classifier can be trained for upcoming cybersecurity incidents with limited data.


\begin{table}[t]
    \centering
    \begin{tabular}{l@{\hspace{1em}}l@{\hspace{1em}}}
    \toprule
    \textbf{Name}                            & \textbf{F1}    \\ \midrule
    Our Approach      & {\scriptsize 80.42/} 80.63(0.27) {\scriptsize /81.07}\\
    → w/o Augmentation      & {\scriptsize 78.48/} 80.33(1.27) {\scriptsize /81.49}\\ 
    → w/o Multi-Level Fine-Tuning    & {\scriptsize 63.95/} 66.16(1.67) {\scriptsize /67.43} \\ 
    → w/o ADAPET             & {\scriptsize 65.33/} 71.33(3.62) {\scriptsize /75.08}                  \\ 
    \bottomrule
    \end{tabular}
    \caption{Detailed evaluation results of the ablation experiments. The values on the left show the minimum, in the middle the mean, in brackets the standard deviation, and on the right the maximum value.}
    \label{tab:ablation_evaluation_results}
\end{table}

\subsubsection{Ablation Studies} \label{ablation}

Finally, we want to give a more detailed insight into our method by showing how each component contributes to the resulting score.
For this purpose, we conducted three further experiments in which we omitted one component in each case and evaluated the other two components.
When multi-level fine-tuning is not used, we evaluate the BERT base model with the auxiliary data of the augmentation method and ADAPET for the learning objective.
Without ADAPET, we train the cybersecurity pre-trained model on the CySecAlert dataset and the final task (with augmented data) with a classifier head. 
In the last experiment, the augmented data are simply omitted, while training the model in the multi-level fine-tuning process with ADAPET.

Upon examination of the results, presented in Table \ref{tab:ablation_evaluation_results}, it becomes clear that leaving out a component worsens the overall results.
The highest loss is reached when the multi-level fine-tuning component is left out, showing how important it is.
This behavior could be due to the many specific cybersecurity words trained by the general language modelling of cybersecurity data (CyBERT) and to fine-tuning by a very related task that already gives the model an idea of how to distinguish between relevant and irrelevant content. 
Furthermore, we can clearly observe that leaving out ADAPET greatly worsens the results.
When compared with the results of the main evaluation presented in Table \ref{tab:evaluation_results}, ADAPET even improves the values significantly more than compared to the baseline.
This shows that ADAPET needs a strong base model to be highly beneficial.
The smallest improvement is made with the augmented data. 
Although the data appeared to be of high quality (see Section \ref{eval:dataaugmentation}), it did not significantly improve the classifier.
Nevertheless, a small increase can be reached and the classifier training got more robust through the additional training data (smallest standard deviation).

\begin{table*}[t]
\centering
\caption{Generated data instances and their most similar original counterparts. The instances created are displayed first and the most similar ones second. URLs are removed from the text.}
\label{table:generated_data}
\begin{tabularx}{\textwidth}{lX}
\toprule
\multirow{10}{*}{\begin{tabular}[c]{@{}c@{}} Positive \end{tabular}} &
  RT If you're running Microsoft Exchange Server on premises, you need to take these urgent security steps now. The zero-day exploits may have already caused a breach of your data. \#infosec \#cybersecurity \#HAFNIUM http://.. \\ \\
  &
  If you are hosting \#MicrosoftExchange on premises you need to take these urgent security steps right now. The zero-day exploits may have already caused a breach of your data. \#infosec \#HAFNIUM http://.. \\ \cmidrule{2-2} 
 &
  RT Black Kingdom ransomware is exploiting the Microsoft Exchange Server ProxyLogon vulnerabilities to encrypt servers. http://.. \\ \\
 &
  Please take Information Security seriously.
  \#CyberAttack can bring your reputation down.
  Another \#ransomware operation known as 'Black Kingdom' is exploiting the Microsoft Exchange Server ProxyLogon vulnerabilities to \#encrypt servers. http://..
   \\ \cmidrule{2-2} 
 &
  RT @SecureList: The ProxyLogon vulnerability in Microsoft Exchange Server is being actively exploited in the wild to install ransomware. http://.. \\ \\
 &
  RT Just as predicted, the Microsoft Exchange exploit chain \#ProxyLogon now confirmed being used to install ransomware \#DEARCRY http://.. \\ \cmidrule{2-2} 
 &
  RT RT @hackerfantastic: Microsoft Exchange Server Remote Code Execution CVE-2021-26855 Exploit.

  \#BugBounty \#RCE \#infosec http://.. \\ \\
 &
  RT Thousands of US companies have been hacked by Chinese hackers using This RCE.

  Microsoft Exchange Server Remote Code Execution CVE-2021-26855 Exploit.

  \#BugBounty \#RCE \#infosec http://.. \\ \cmidrule{2-2} 
 &
  RT @ryan\_a\_h: Microsoft just released their quarterly updates which include a patch for the Exchange zero-day. You can find more information here: http://.. \\ \\
 &
  If you are hosting \#MicrosoftExchange on premises you need to take these urgent security steps right now. The zero-day exploits may have already caused a breach of your data. \#infosec \#HAFNIUM http://.. \\ \cmidrule{1-2} 
\multirow{10}{*}{Negative} &
  Microsoft Exchange Server Attack Escalation Prompts \#Patching Panic

  \#cybersecurity \#vulnerabilities http://.. \\ \\
 &
  \#MicrosoftExchange Server Attack Escalation Prompts \#Patching Panic

  \#cybersecurity \#vulnerabilities http://.. \\ \cmidrule{2-2} 
 &
 \#Technology \#Cybersecurity Microsoft Exchange Hackers Also Breached European Banking Authority \#AiUpNow \#techy http://.. \\ \\
 &
 \#Technology \#TechNews Microsoft Exchange Hackers Also Breached European Banking Authority \#Cybersecurity \#AiUpNow \#techy via http://.. \\ \cmidrule{2-2}
 &
 RT Microsoft Exchange Server has been hacked – here’s what you need to know http://.. \\ \\
 &
  RT Here's what we know so far about the massive Microsoft Exchange hack http://.. \\ \cmidrule{2-2} 
 &
  Microsoft Exchange Server Flaws Expose Millions of Emails to Attack http://.. \\ \\
 &
  RT Here's what we know so far about the massive Microsoft Exchange hack http://.. \\ \cmidrule{2-2} 
 &
  Protected: Microsoft Exchange Server Attacks Escalate to Government, Healthcare and Financial Institutions http://.. \\ \\
 &
  The Microsoft Exchange hacks: How they started and where we are http://.. \\ 
  \bottomrule
\end{tabularx}
\end{table*}
\section{Conclusion and Discussion} \label{discussion}

CTI, the collection of evidence-based knowledge of cybersecurity threats, is highly relevant for identifying and remediating security incidents. 
Professionals, security providers, CERTs, as well as many others in the cybersecurity realm can gain important information about the incidents, such as how severe they may be, which software and systems are affected, how to be protected, and if exploits exist.
The challenges lie in the information overload and the high dynamics associated with every new threat event. 
To our knowledge, this is the first work to address this issue by proposing a framework for specialized CTI. 
It consists of several components that allow the end user to label only a few data instances (tested here with 32 instances) to obtain a classifier that is comparable to one trained with 1800 instances.
We also constructed a dataset labeled by three cybersecurity experts showing that this method indeed overcomes the problem of information overload and addresses high dynamics by being easily adaptable to new incidents.

\subsection{Practical, Theoretical, and Empirical Contributions} \label{contributions}

Considering our findings, the study revealed (P) practical, (T) theoretical, and (E) empirical contributions:

\textbf{(P) A novel pipeline for detecting specialized cyber threat information.} 
Our work provides an approach to the detection of specific cyber threat information that is aligned with the circumstances of such events. 
These circumstances include that information has to be gathered fast in the early stages of the events and that security institutions and experts do not have the time and capacity to label many instances. 
Therefore, we combine few-shot learning with multi-level fine-tuning and data augmentation to produce classifiers that only need few instances to perform with high quality. 
For few-shot learning we utilize ADAPET by \citet{tamImprovingSimplifyingPattern2021} combined with the multi-level fine-tuning process. 
For data augmentation we use GPT-3 to create instances with novel linguistic patterns. 
Our pipeline reaches a F1-score of 80.63 on a specialized cyber threat dataset, which is 21.93 points above the score of a classical learning scheme. Other work, such as the cyber threat detection systems of \cite{riebeCySecAlertAlertGeneration2021} or \cite{Sceller2017}, allow for coarse-grained information gathering.
To the best of our knowledge, our system is the first to provide rapid detection of specialized cyber threat information.

\textbf{(T) New few-shot learning technique based on multi-level fine-tuning.} 
We propose a novel few-shot learning approach for creating classifiers of high quality with a smaller amount of training data.
The idea behind this approach is to fine-tune a machine learning model in several levels where enough data is available (see Figure \ref{fig:mulitlevel}). 
In our study we first further trained a BERT model on a general cybersecurity corpus. 
This model was then trained on a general Twitter cybersecurity relevance dataset. 
From this point, the model has a fundamental understanding of cybersecurity texts and is also able to distinguish cybersecurity-related content from irrelevant content. 
With this pre-trained knowledge, the model only needs few data instances to be able to differentiate specific cybersecurity content. 
As shown in this study, this new technique can also be combined with other techniques like ADAPET or data augmentation to further reduce the amount of needed training data.
However, we show that this multi-stage fine-tuning approach has the greatest impact on classification quality of all techniques (+14.47 F1, see Table \ref{tab:ablation_evaluation_results}).
The multi-level fine-tuning approach significantly advances research in few-shot learning, as it allows for a much higher model quality and at the same time can be combined with previous few-shot studies, such as ADAPET \citep{tamImprovingSimplifyingPattern2021}, DART \citep{zhangDifferentiablePromptMakes2022}, or PERFECT \citep{mahabadiPERFECTPromptfreeEfficient2022}.

\textbf{(T) New insights on data augmentation with large pre-trained language models.} 
In our study, we also implemented a data augmentation technique that combines the works of \citet{yooGPT3MixLeveragingLargescale2021} and \citet{bayerDataAugmentationNatural2021}.
As in the former, we used the large language model GPT-3 with a prompting strategy and filtered the generated instances with a human-in-the-loop technique, as in the latter.
The idea is that GPT-3 can create instances with a very high degree of novelty, resulting in some very valuable instances. 
However, this novelty comes with the problem of poor label preservation, as the instances may be too far away from the class.
For this reason, we also introduced this filtering strategy where the original labeled data of a class is compared with the generated data and those that are too far away from the original data are discarded. 
The boundary is determined by an expert who examines those instances close to a predefined boundary.
As shown in section \ref{eval:dataaugmentation} and table \ref{table:generated_data}, this procedure generates instances with very different transformation patterns, including word substitution, paraphrasing, and partial removal. 
It even leads to instances that are entirely novel.
However, in section \ref{ablation}, we showed that omitting this method from the overall pipeline only slightly reduces the resulting score.
This means that the model learns very little from the augmented data when multi-level fine-tuning and ADAPET are already used.
Nevertheless, the evaluation results show a reduction in the standard deviation, which shows that the model has become more robust with the artificial data.

\textbf{(E) A specialized CTI dataset for further research purposes.} 
In this study we created a CTI dataset based on the 2021 Microsoft Exchange Server data breach. 
The dataset was constructed by three experts. 
The guidelines have been revised several times in an attempt to flesh out the concept of cyber threat analysis as much as possible.
Along with the code and the dataset, the guidelines are available in the repository.
All annotators reached a good intercoder reliability showing that the guidelines and the general annotation process was successful. 
Further research can benefit from this dataset as it is, to our knowledge, the first to contain a relevance coding regarding CTI in Twitter in relation to a specific cybersecurity event.

\subsection{Limitations and Outlook} \label{limitations}

In terms of the overall concept, we look forward to research studies testing the performance of this approach in other domains.
For example, it would be interesting to see if the same improvements can be achieved in medical or crisis domains, where data is also scarce.
Moreover, our experiments are limited to the BERT base model. 
It would be interesting to see if the improvements are as high when a larger model like RoBERTa \citep{liuRoBERTaRobustlyOptimized2019} is used.
Likewise, one could also test other language models for the data augmentation technique.
Especially interesting would be to test if open source models, like GPT-NeoX-20B \citep{blackGPTNeoX20BOpenSourceAutoregressive2022}, reach a good augmentation performance.

A part of our experiments was to fine-tune the model on the CySecAlert dataset of \citet{riebeCySecAlertAlertGeneration2021}.
The authors of this work propose an active learning component to achieve high classification scores with less data. 
With a view to future research, it might be sensible to also include active learning into the concept of our approach to further increase the classification quality.
In practice, our approach would in the worst case lead to users labelling very similar examples, resulting in poor execution of data augmentation and poor classification quality, which can happen quickly when labelling such a small amount of data. 
Therefore, an active learning system could help to collect very different examples. 
Otherwise, experts can also be trained to label diverse examples. 

\begin{acknowledgements}
This work has been co-funded by the German Federal Ministry of Education and Research (BMBF) in the project CYWARN (13N15407) and funded by the Deutsche Forschungsgemeinschaft (DFG, German Research Foundation) – SFB 1119 (CROSSING) – 236615297, as well as the German Federal Ministry of Education and Research and the Hessian Ministry of Higher Education, Research, Science and the Arts within their joint support of the National Research Center for Applied Cybersecurity ATHENE. Calculations for this research were conducted on the Lichtenberg high performance computer of the TU Darmstadt. 
\end{acknowledgements}

%
%

\bibliographystyle{spbasic}      
\bibliography{references}   

\begin{thebibliography}{42}
\providecommand{\natexlab}[1]{#1}
\providecommand{\url}[1]{{#1}}
\providecommand{\urlprefix}{URL }
\expandafter\ifx\csname urlstyle\endcsname\relax
  \providecommand{\doi}[1]{DOI~\discretionary{}{}{}#1}\else
  \providecommand{\doi}{DOI~\discretionary{}{}{}\begingroup
  \urlstyle{rm}\Url}\fi
\providecommand{\eprint}[2][]{\url{#2}}

\bibitem[{Abu et~al.(2018)Abu, Selamat, Ariffin, and Yusof}]{abu2018cyber}
Abu MS, Selamat SR, Ariffin A, Yusof R (2018) Cyber threat intelligence--issue
  and challenges. Indonesian Journal of Electrical Engineering and Computer
  Science 10(1):371--379

\bibitem[{Anaby-Tavor et~al.(2020)Anaby-Tavor, Carmeli, Goldbraich, Kantor,
  Kour, Shlomov, Tepper, and Zwerdling}]{Anaby-Tavor2020}
Anaby-Tavor A, Carmeli B, Goldbraich E, Kantor A, Kour G, Shlomov S, Tepper N,
  Zwerdling N (2020) Do not have enough data? {Deep} learning to the rescue!
  Proceedings of the AAAI \urlprefix\url{http://arxiv.org/abs/1911.03118}

\bibitem[{Bayer et~al.(2021)Bayer, Kaufhold, Buchhold, Keller, Dallmeyer, and
  Reuter}]{bayerDataAugmentationNatural2021}
Bayer M, Kaufhold MA, Buchhold B, Keller M, Dallmeyer J, Reuter C (2021) Data
  {Augmentation} in {Natural} {Language} {Processing}: {A} {Novel} {Text}
  {Generation} {Approach} for {Long} and {Short} {Text} {Classifiers}.
  International Journal of Machine Learning and Cybernetics (IJMLC)
  \doi{10.1007/s13042-022-01553-3},
  \urlprefix\url{http://arxiv.org/abs/2103.14453}

\bibitem[{Bayer et~al.(2022)Bayer, Kaufhold, and
  Reuter}]{bayerSurveyDataAugmentation2022}
Bayer M, Kaufhold MA, Reuter C (2022) A {Survey} on {Data} {Augmentation} for
  {Text} {Classification}. ACM Computing Surveys p 3544558,
  \doi{10.1145/3544558}, \urlprefix\url{https://dl.acm.org/doi/10.1145/3544558}

\bibitem[{Belinkov and Bisk(2018)}]{Belinkov2018}
Belinkov Y, Bisk Y (2018) Synthetic and natural noise both break neural machine
  translation. In: Proceedings of {ICLR}

\bibitem[{Beltagy et~al.(2019)Beltagy, Lo, and
  Cohan}]{beltagySciBERTPretrainedLanguage2019}
Beltagy I, Lo K, Cohan A (2019) {SciBERT}: {A} {Pretrained} {Language} {Model}
  for {Scientific} {Text}. \urlprefix\url{http://arxiv.org/abs/1903.10676},
  arXiv:1903.10676 [cs]

\bibitem[{Black et~al.(2022)Black, Biderman, Hallahan, Anthony, Gao, Golding,
  He, Leahy, McDonell, Phang, Pieler, Prashanth, Purohit, Reynolds, Tow, Wang,
  and Weinbach}]{blackGPTNeoX20BOpenSourceAutoregressive2022}
Black S, Biderman S, Hallahan E, Anthony Q, Gao L, Golding L, He H, Leahy C,
  McDonell K, Phang J, Pieler M, Prashanth US, Purohit S, Reynolds L, Tow J,
  Wang B, Weinbach S (2022) {GPT}-{NeoX}-{20B}: {An} {Open}-{Source}
  {Autoregressive} {Language} {Model}.
  \urlprefix\url{http://arxiv.org/abs/2204.06745}

\bibitem[{Bragg et~al.(2021)Bragg, Cohan, Lo, and
  Beltagy}]{braggFLEXUnifyingEvaluation2021}
Bragg J, Cohan A, Lo K, Beltagy I (2021) {FLEX}: {Unifying} {Evaluation} for
  {Few}-{Shot} {NLP}. arXiv p~14

\bibitem[{Brown et~al.(2020)Brown, Mann, Ryder, Subbiah, Kaplan, Dhariwal,
  Neelakantan, Shyam, Sastry, Askell, Agarwal, Herbert-Voss, Krueger, Henighan,
  Child, Ramesh, Ziegler, Wu, Winter, Hesse, Chen, Sigler, Litwin, Gray, Chess,
  Clark, Berner, McCandlish, Radford, Sutskever, and Amodei}]{Brown2020}
Brown TB, Mann B, Ryder N, Subbiah M, Kaplan J, Dhariwal P, Neelakantan A,
  Shyam P, Sastry G, Askell A, Agarwal S, Herbert-Voss A, Krueger G, Henighan
  T, Child R, Ramesh A, Ziegler DM, Wu J, Winter C, Hesse C, Chen M, Sigler E,
  Litwin M, Gray S, Chess B, Clark J, Berner C, McCandlish S, Radford A,
  Sutskever I, Amodei D (2020) Language models are few-shot learners. In:
  {NeurIPS}, \urlprefix\url{http://arxiv.org/abs/2005.14165}

\bibitem[{Chatterjee and Thekdi(2020)}]{CHATTERJEE2020106664}
Chatterjee S, Thekdi S (2020) An iterative learning and inference approach to
  managing dynamic cyber vulnerabilities of complex systems. Reliability
  Engineering \& System Safety 193:106664,
  \doi{https://doi.org/10.1016/j.ress.2019.106664},
  \urlprefix\url{https://www.sciencedirect.com/science/article/pii/S0951832018314558}

\bibitem[{Devlin et~al.(2018)Devlin, Chang, Lee, and Toutanova}]{Devlin2018}
Devlin J, Chang MW, Lee K, Toutanova K (2018) {BERT}: {Pre}-training of deep
  bidirectional transformers for language understanding (Mlm),
  \urlprefix\url{http://arxiv.org/abs/1810.04805}

\bibitem[{Dionísio et~al.(2020)Dionísio, Alves, Ferreira, and
  Bessani}]{dionisioEndtoendCyberthreatDetection2020}
Dionísio N, Alves F, Ferreira PM, Bessani A (2020) Towards end-to-end
  {Cyberthreat} {Detection} from {Twitter} using {Multi}-{Task} {Learning}. In:
  2020 {International} {Joint} {Conference} on {Neural} {Networks} ({IJCNN}),
  pp 1--8, \doi{10.1109/IJCNN48605.2020.9207159}, iSSN: 2161-4407

\bibitem[{Fang et~al.(2020)Fang, Gao, Liu, and
  Huang}]{fangDetectingCyberThreat2020a}
Fang Y, Gao J, Liu Z, Huang C (2020) Detecting {Cyber} {Threat} {Event} from
  {Twitter} {Using} {IDCNN} and {BiLSTM}. Applied Sciences 10(17):5922,
  \doi{10.3390/app10175922},
  \urlprefix\url{https://www.mdpi.com/2076-3417/10/17/5922}

\bibitem[{Gao et~al.(2021)Gao, Fisch, and
  Chen}]{gaoMakingPretrainedLanguage2021}
Gao T, Fisch A, Chen D (2021) Making {Pre}-trained {Language} {Models} {Better}
  {Few}-shot {Learners}. \urlprefix\url{http://arxiv.org/abs/2012.15723}

\bibitem[{Jiang et~al.(2020)Jiang, He, Chen, Liu, Gao, and
  Zhao}]{jiangSMARTRobustEfficient2020}
Jiang H, He P, Chen W, Liu X, Gao J, Zhao T (2020) {SMART}: {Robust} and
  {Efficient} {Fine}-{Tuning} for {Pre}-trained {Natural} {Language} {Models}
  through {Principled} {Regularized} {Optimization}. In: Proceedings of the
  58th {Annual} {Meeting} of the {Association} for {Computational}
  {Linguistics}, Association for Computational Linguistics, Online, pp
  2177--2190, \doi{10.18653/v1/2020.acl-main.197},
  \urlprefix\url{https://www.aclweb.org/anthology/2020.acl-main.197}

\bibitem[{Kaufhold et~al.(2022)Kaufhold, Basyurt, Eyilmez, Ag, Stöttinger,
  Reuter, and Sercan}]{kaufholdCyberThreatObservatory2022}
Kaufhold MA, Basyurt AS, Eyilmez K, Ag V, Stöttinger M, Reuter C, Sercan A
  (2022) Cyber {Threat} {Observatory}: {Design} and {Evaluation} of an
  {Interactive} {Dashboard} for {Computer} {Emergency} {Response} {Teams}. ECIS
  2022 p~18

\bibitem[{Lan et~al.(2020)Lan, Chen, Goodman, Gimpel, Sharma, and
  Soricut}]{lanALBERTLiteBERT2020}
Lan Z, Chen M, Goodman S, Gimpel K, Sharma P, Soricut R (2020) {ALBERT}: {A}
  {Lite} {BERT} for {Self}-supervised {Learning} of {Language}
  {Representations}. \urlprefix\url{http://arxiv.org/abs/1909.11942},
  arXiv:1909.11942 [cs]

\bibitem[{Le~Sceller et~al.(2017)Le~Sceller, Karbab, Debbabi, and
  Iqbal}]{Sceller2017}
Le~Sceller Q, Karbab EB, Debbabi M, Iqbal F (2017) Sonar: Automatic detection
  of cyber security events over the twitter stream. In: Proceedings of the 12th
  International Conference on Availability, Reliability and Security,
  Association for Computing Machinery, New York, NY, USA, ARES '17,
  \doi{10.1145/3098954.3098992},
  \urlprefix\url{https://doi.org/10.1145/3098954.3098992}

\bibitem[{Lee et~al.(2019)Lee, Yoon, Kim, Kim, Kim, So, and
  Kang}]{leeBioBERTPretrainedBiomedical2019}
Lee J, Yoon W, Kim S, Kim D, Kim S, So CH, Kang J (2019) {BioBERT}: a
  pre-trained biomedical language representation model for biomedical text
  mining. Bioinformatics p btz682, \doi{10.1093/bioinformatics/btz682},
  \urlprefix\url{https://academic.oup.com/bioinformatics/advance-article/doi/10.1093/bioinformatics/btz682/5566506}

\bibitem[{Liu et~al.(2019)Liu, Ott, Goyal, Du, Joshi, Chen, Levy, Lewis,
  Zettlemoyer, Stoyanov, and Allen}]{liuRoBERTaRobustlyOptimized2019}
Liu Y, Ott M, Goyal N, Du J, Joshi M, Chen D, Levy O, Lewis M, Zettlemoyer L,
  Stoyanov V, Allen PG (2019) {RoBERTa}: {A} {Robustly} {Optimized} {BERT}
  {Pretraining} {Approach}. Tech. rep.,
  \urlprefix\url{https://github.com/pytorch/fairseq}

\bibitem[{Longpre et~al.(2020)Longpre, Wang, and DuBois}]{Longpre2020}
Longpre S, Wang Y, DuBois C (2020) How effective is task-agnostic data
  augmentation for pretrained transformers? In: Findings of {EMNLP}

\bibitem[{Mahabadi et~al.(2022)Mahabadi, Zettlemoyer, Henderson, Saeidi,
  Mathias, Stoyanov, and Yazdani}]{mahabadiPERFECTPromptfreeEfficient2022}
Mahabadi RK, Zettlemoyer L, Henderson J, Saeidi M, Mathias L, Stoyanov V,
  Yazdani M (2022) {PERFECT}: {Prompt}-free and {Efficient} {Few}-shot
  {Learning} with {Language} {Models}.
  \urlprefix\url{http://arxiv.org/abs/2204.01172}, arXiv:2204.01172 [cs]

\bibitem[{Martin et~al.(2020)Martin, Muller, Ortiz~Su{\'a}rez, Dupont, Romary,
  de~la Clergerie, Seddah, and Sagot}]{martin-etal-2020-camembert}
Martin L, Muller B, Ortiz~Su{\'a}rez PJ, Dupont Y, Romary L, de~la Clergerie
  {\'E}, Seddah D, Sagot B (2020) {C}amem{BERT}: a tasty {F}rench language
  model. In: Proceedings of the 58th Annual Meeting of the Association for
  Computational Linguistics, Association for Computational Linguistics, Online,
  pp 7203--7219,
  \urlprefix\url{https://www.aclweb.org/anthology/2020.acl-main.645}

\bibitem[{McMillan(2013)}]{robmcmillanDefinitionThreatIntelligence2013}
McMillan R (2013) Definition: {Threat} {Intelligence}.
  \urlprefix\url{https://www.gartner.com/en/documents/2487216}

\bibitem[{Mittal et~al.(2016)Mittal, Das, Mulwad, Joshi, and
  Finin}]{mittal2016cybertwitter}
Mittal S, Das PK, Mulwad V, Joshi A, Finin T (2016) Cybertwitter: Using twitter
  to generate alerts for cybersecurity threats and vulnerabilities. In: 2016
  IEEE/ACM International Conference on Advances in Social Networks Analysis and
  Mining (ASONAM), IEEE, pp 860--867

\bibitem[{Mosolova et~al.(2018)Mosolova, Fomin, and
  Bondarenko}]{mosolovaTextAugmentationNeural2018}
Mosolova AV, Fomin VV, Bondarenko IY (2018) Text augmentation for neural
  networks. CEUR Workshop Proceedings 2268:104--109

\bibitem[{Pan(2020)}]{pan2020transfer}
Pan SJ (2020) Transfer learning. Learning 21:1--2

\bibitem[{Queiroz~Abonizio and
  Barbon~Junior(2020)}]{queirozabonizioPretrainedDataAugmentation2020}
Queiroz~Abonizio H, Barbon~Junior S (2020) Pre-trained {Data} {Augmentation}
  for {Text} {Classification}. In: Lecture {Notes} in {Computer} {Science}
  (including subseries {Lecture} {Notes} in {Artificial} {Intelligence} and
  {Lecture} {Notes} in {Bioinformatics}), Springer Science and Business Media
  Deutschland GmbH, vol 12319 LNAI, pp 551--565,
  \doi{10.1007/978-3-030-61377-8_38}, iSSN: 16113349

\bibitem[{Reimers and
  Gurevych(2019)}]{reimersSentenceBERTSentenceEmbeddings2019}
Reimers N, Gurevych I (2019) Sentence-{BERT}: {Sentence} {Embeddings} using
  {Siamese} {BERT}-{Networks}. \doi{10.18653/v1/d19-1410}

\bibitem[{Riebe et~al.(2021{\natexlab{a}})Riebe, Kaufhold, and
  Reuter}]{riebeImpactOrganizationalStructure2021}
Riebe T, Kaufhold MA, Reuter C (2021{\natexlab{a}}) The {Impact} of
  {Organizational} {Structure} and {Technology} {Use} on {Collaborative}
  {Practices} in {Computer} {Emergency} {Response} {Teams}: {An} {Empirical}
  {Study}. Proceedings of the ACM on Human-Computer Interaction 5(CSCW2):1--30,
  \doi{10.1145/3479865}, \urlprefix\url{https://dl.acm.org/doi/10.1145/3479865}

\bibitem[{Riebe et~al.(2021{\natexlab{b}})Riebe, Wirth, Bayer, Kühn, Kaufhold,
  Knauthe, Guthe, and Reuter}]{riebeCySecAlertAlertGeneration2021}
Riebe T, Wirth T, Bayer M, Kühn P, Kaufhold MA, Knauthe V, Guthe S, Reuter C
  (2021{\natexlab{b}}) {CySecAlert}: {An} {Alert} {Generation} {System} for
  {Cyber} {Security} {Events} {Using} {Open} {Source} {Intelligence} {Data}.
  In: Gao D, Li Q, Guan X, Liao X (eds) Information and {Communications}
  {Security}, Springer International Publishing, Cham, Lecture {Notes} in
  {Computer} {Science}, pp 429--446, \doi{10.1007/978-3-030-86890-1_24}

\bibitem[{Rodriguez and Okamura(2019)}]{rodriguez2019generating}
Rodriguez A, Okamura K (2019) Generating real time cyber situational awareness
  information through social media data mining. In: 2019 IEEE 43rd annual
  computer software and applications conference (COMPSAC), IEEE, vol~2, pp
  502--507

\bibitem[{Sabottke et~al.(2015)Sabottke, Suciu, and Dumitras}]{Sabottke2015}
Sabottke C, Suciu O, Dumitras T (2015) Vulnerability disclosure in the age of
  social media: Exploiting twitter for predicting {Real-World} exploits. In:
  24th USENIX Security Symposium (USENIX Security 15), USENIX Association,
  Washington, D.C., pp 1041--1056,
  \urlprefix\url{https://www.usenix.org/conference/usenixsecurity15/technical-sessions/presentation/sabottke}

\bibitem[{Schick and Schütze(2021)}]{schickExploitingClozeQuestions2021}
Schick T, Schütze H (2021) Exploiting {Cloze} {Questions} for {Few} {Shot}
  {Text} {Classification} and {Natural} {Language} {Inference}.
  \urlprefix\url{http://arxiv.org/abs/2001.07676}, arXiv:2001.07676 [cs]

\bibitem[{Sun et~al.(2020)Sun, Xia, Yin, Liang, Yu, and
  He}]{sunMixuptransfomerDynamicData2020}
Sun L, Xia C, Yin W, Liang T, Yu PS, He L (2020) Mixup-transfomer: {Dynamic}
  data augmentation for {NLP} tasks. \doi{10.18653/v1/2020.coling-main.305},
  iSSN: 23318422 Publication Title: arXiv \_eprint: 2010.02394

\bibitem[{Tam et~al.(2021)Tam, Menon, Bansal, Srivastava, and
  Raffel}]{tamImprovingSimplifyingPattern2021}
Tam D, Menon RR, Bansal M, Srivastava S, Raffel C (2021) Improving and
  {Simplifying} {Pattern} {Exploiting} {Training}.
  \urlprefix\url{http://arxiv.org/abs/2103.11955}, number: arXiv:2103.11955
  arXiv:2103.11955 [cs]

\bibitem[{Taylor(1953)}]{taylorClozeProcedureNew1953}
Taylor WL (1953) “{Cloze} {Procedure}”: {A} {New} {Tool} for {Measuring}
  {Readability}. Journalism Quarterly 30(4):415--433,
  \doi{10.1177/107769905303000401},
  \urlprefix\url{http://journals.sagepub.com/doi/10.1177/107769905303000401}

\bibitem[{Torrey and Shavlik(2010)}]{torrey2010transfer}
Torrey L, Shavlik J (2010) Transfer learning. In: Handbook of research on
  machine learning applications and trends: algorithms, methods, and
  techniques, IGI global, pp 242--264

\bibitem[{Tounsi and Rais(2018)}]{TOUNSI2018212}
Tounsi W, Rais H (2018) A survey on technical threat intelligence in the age of
  sophisticated cyber attacks. Computers \& Security 72:212--233,
  \doi{https://doi.org/10.1016/j.cose.2017.09.001},
  \urlprefix\url{https://www.sciencedirect.com/science/article/pii/S0167404817301839}

\bibitem[{Wagner et~al.(2019)Wagner, Mahbub, Palomar, and
  Abdallah}]{WAGNER2019101589}
Wagner TD, Mahbub K, Palomar E, Abdallah AE (2019) Cyber threat intelligence
  sharing: Survey and research directions. Computers \& Security 87:101589,
  \doi{https://doi.org/10.1016/j.cose.2019.101589},
  \urlprefix\url{https://www.sciencedirect.com/science/article/pii/S016740481830467X}

\bibitem[{Yoo et~al.(2021)Yoo, Park, Kang, Lee, and
  Park}]{yooGPT3MixLeveragingLargescale2021}
Yoo KM, Park D, Kang J, Lee SW, Park W (2021) {GPT3Mix}: {Leveraging}
  {Large}-scale {Language} {Models} for {Text} {Augmentation}. In: Findings of
  the {Association} for {Computational} {Linguistics}: {EMNLP} 2021,
  Association for Computational Linguistics, Punta Cana, Dominican Republic, pp
  2225--2239, \doi{10.18653/v1/2021.findings-emnlp.192},
  \urlprefix\url{https://aclanthology.org/2021.findings-emnlp.192}

\bibitem[{Zhang et~al.(2022)Zhang, Li, Chen, Deng, Bi, Tan, Huang, and
  Chen}]{zhangDifferentiablePromptMakes2022}
Zhang N, Li L, Chen X, Deng S, Bi Z, Tan C, Huang F, Chen H (2022)
  Differentiable {Prompt} {Makes} {Pre}-trained {Language} {Models} {Better}
  {Few}-shot {Learners}. \urlprefix\url{http://arxiv.org/abs/2108.13161},
  arXiv:2108.13161 [cs]

\end{thebibliography}

%
%


\end{document}